\documentclass[preprint]{revtex4}

\usepackage{amsmath}
\usepackage{amssymb}
\usepackage{graphicx}
\usepackage{psfrag}
\usepackage{bm}
\usepackage{sidecap}
\usepackage{listings}  % For inserting code 

\begin{document}

%\doublespacing

\title{Correlation functions of integrable models:  a description of the ABACUS algorithm}

\author{Jean-S\'ebastien Caux\\
%}
%\address{
Institute for Theoretical Physics, Universiteit van Amsterdam, \\
Valckenierstraat 65,
1018 XE Amsterdam, The Netherlands, \\
{\sf J.S.Caux@uva.nl}}

\begin{abstract}
Recent developments in the theory of integrable models have provided the means of
calculating dynamical correlation functions of some important observables in systems
such as Heisenberg spin chains and one-dimensional atomic gases.  
This article explicitly describes how such calculations are generally implemented in
the ABACUS {\sf C++} library, emphasizing the universality in treatment of different cases 
coming as a consequence of unifying features within the Bethe Ansatz.  
\end{abstract}

\maketitle

\section{Introduction}
\label{sec:Introduction}
The Bethe Ansatz, introduced in 1931 \cite{BetheZP71} (very shortly after the appearance of quantum mechanics) is often
viewed as an arcane subject, manifestly interesting for mathematically-minded theorists but not for a broad audience.  At least 
two good reasons for this can be given.  First, the subject of integrability is rather specialized and demanding in
itself \cite{GaudinBOOK,KorepinBOOK,TakahashiBOOK,SutherlandBOOK}:
it is after all made up of a collection of somewhat elaborate methods having only limited general applicability,
and is therefore not typically considered worthy of inclusion in the standard condensed matter curriculum 
(the fact remains that the technology developed around integrable models has proven its usefulness in many ways, perhaps most importantly
in providing many contributions to our understanding of strongly-correlated physics in one dimension \cite{GiamarchiBOOK},
but also in providing reliable beacons for the testing and benchmarking of more widely-used field-theory- or numerics-based 
methods for such systems).
Second, and perhaps more importantly, is that the experimental relevance of integrable systems
remains limited.  Good realizations of one-dimensional integrable quantum systems do exit,
but are the exception rather than the rule;  when a good realization is found, integrability
often cannot provide quantitative predictions for the most important observable quantities.

This last difficulty is related to the long-standing inability of the Bethe Ansatz to provide
results for even simple correlation functions of local physical observables.  The source of this 
difficulty lies in the admittedly unwieldy form of Bethe wavefunctions, which are composed of a
factorially large sum of free waves with nontrivial relative amplitudes.  The action of local
operators on these wavefunctions is difficult to write down, and since this forms the starting
point in computing any correlation function, progress has been tedious, slow and limited.

Recent years have however seen some important developments occur at a rapid pace in the calculation of 
dynamical correlations for finite systems at zero temperature.  This stems from the appearance of a number
of important results in the field of integrability.  First and foremost is Slavnov's theorem on scalar products,
providing initial results on form factors and correlation functions \cite{SlavnovTMP79,SlavnovTMP82}.  
The general problem of describing the action of local operators on Bethe wavefunctions was then resolved 
with the breakthrough solution of the quantum inverse problem for spin chains
\cite{KitanineNPB554,KitanineNPB567}, opening the door to the computation of general correlation functions of the $XXZ$ chain.  
The problem of obtaining correlation functions from these representations
was then initiated in \cite{BiegelEL59,BiegelJPA36,SatoJPSJ73}, with a study of traces over two-particle intermediate states.
The ABACUS algorithm was thereafter developed in order to tackle the summation of all important 
general multiparticle excitations in Heisenberg spin chains and the one-dimensional Bose gas
\cite{CauxPRL95,CauxJSTAT05P09003,CauxPRA74,CauxJSTAT07P01008}.  
The present article only provides a `conceptual-level' description of the features and
functioning of ABACUS, only emphasizing unifying features of this class of calculations without going deeply into specific examples.
My main objective in this article is to give a more or less detailed discussion of the strategies adopted in ABACUS,
to define some terminology which will prove useful in future works, and to make some observations on the results obtained.
A separate user guide \cite{ABACUS_Users_Guide} will provide implementation-level details of the library and
its usage, including important primary class and function declarations.  
Extensive and systematic results for various models will form the subject of specialized publications.  
The subject of integrability itself will be covered in a separate set of lecture notes \cite{Bethe_Ansatz_Course_JSC}.  
The general field of correlation functions of strongly-correlated one-dimensional quantum models is the subject
of an upcoming review \cite{CauxADVPHYS}.

\section{Defining the problem}
\label{sec:Defining}
Let us for definiteness consider a quantum-mechanical model on a lattice, with Hamiltonian
$H$ which we assume for convenience (and without true loss of generality) to act in a Hilbert
space formed by the tensoring of finitely many single-site local Hilbert spaces.  Letting
$N$ be the number of such sites and using $j = 1, ..., N$ as a site label, all operators can
be reconstructed from the self-dual set ${\cal O}^a_j$ of local physical operators,
where $a$ is some on-site representation index.  

The general problem which we will address is the calculation of two-point zero temperature equilibrium
correlation functions of the form
\begin{eqnarray}
\langle {\cal O}^a_j (t) {\cal O}^{\bar{a}}_{j'} (0) \rangle,
\label{eq:OO}
\end{eqnarray}
where $\langle ... \rangle$ represents the ground-state expectation value and 
${\cal O}^{\bar a}_j = ({\cal O}^a_j)^{\dagger}$, for arbitrary time differences $t$ and
distances $j - j'$.  Equivalently, we can consider the Fourier transform of this correlation function,
which (by a slight abuse of terminology) will be called the dynamical structure factor (DSF):
\begin{eqnarray}
S^{a \bar{a}} (k, \omega) = \frac{1}{N} \sum_{j, j' = 1}^N e^{-ik(j-j')} \int_{-\infty}^{\infty} dt ~e^{i\omega t}
\langle {\cal O}^a_j (t) {\cal O}^{\bar{a}}_{j'} (0) \rangle.
\label{eq:DSF}
\end{eqnarray}
A complete knowledge of the DSF for all momenta and frequencies naturally allows to reconstruct (\ref{eq:OO})
without approximation.  In reality, however, (\ref{eq:DSF}) itself is of more direct usefulness since
it is often the quantity which can be related to energy- and momentum-resolved experimental measurements
in linear response.

Calculating the DSF for interacting models (by which is meant models where multiparticle states are not created
by simple products of single-particle creation operators) is a very complex task.  
The product of quantum operators acting at different times and places is best dealt with
by introducing a summation over intermediate eigenstates (labeled by greek indices), allowing to resolve the time
dependence explicitly.
Using the space Fourier transform ${\cal O}_k^a = \sum_j e^{-ikj} {\cal O}_j^a$, summing over
lattice sites and performing the time integration leads to 
the Lehmann series representation
\begin{eqnarray}
S^{a\bar{a}} (k, \omega) = \frac{2\pi}{N} \sum_{\mu} | \langle \lambda^0 | {\cal O}_k^a | \mu \rangle |^2
%\delta_{q, q_{\lambda}- q_0} 
\delta (\omega - E_{\mu} + E_0)
\label{zero_T_DSF}
\end{eqnarray}
where $E_0$ 
is the energy 
of the ground state $|\lambda^0\rangle$.  The matrix elements $\langle \lambda | {\cal O} | \mu \rangle$ 
of local operators in the eigenstates basis are also known as form factors
(for continuum models);  their norm square is also often called transition rate in
the literature, in view of the correspondence of (\ref{zero_T_DSF}) to Fermi's
golden rule.

We can make a simple `wish list' of the elements needed to provide 
a computation of such dynamical structure factors:
\begin{enumerate}
\item an orthonormal eigenstate basis;
\item form factors (matrix elements) $\langle \lambda | {\cal O}_k^a | \mu \rangle$ of operators ${\cal O}_k^a$ in this basis;
\item a way to perform the summation over intermediate states.  
\end{enumerate}
In general, this is an immensely complicated problem, which can typically
be performed for models which correspond or can be mapped to free particles, 
or which benefit from a conformal field theory description.  
Simplistic treatments of interacting systems in general do not provide us with any of the three needed
elements.

In one dimension, however, the technology of integrability does provide us with
some of the three elements above, for a number of interesting models.  The
Bethe Ansatz provides the eigenstates basis, and the Algebraic Bethe Ansatz
(together with the solution of the quantum inverse problem) provides us with
form factors.  The summation over intermediate states remains however an open
problem, for which partial analytical results are possible in some restricted circumstances, 
but which still demands more often than not a numerical solution.

The ABACUS library has recently been developed in order to perform the
calculation of dynamical structure factors of the known integrable models for
which eigenstates and form factors can be explicitly constructed.  ABACUS
(\underline{A}lgebraic \underline{B}ethe \underline{A}nsatz-based \underline{C}omputation of \underline{U}niversal \underline{S}tructure factors)
provides functions for representing and constructing eigenstates, calculating
form factors of a number of physical operators of interest, but also implements
a scanning algorithm over intermediate states allowing to construct the 
correlations explicitly.  My main motivations in developing ABACUS were primarily
to provide quantitative predictions first for experiments in low-dimensional magnetism
and later on cold atomic gases, but also to provide results allowing cross-checking
of more generally applicable alternative theoretical (analytical or numerical) 
approaches to the dynamics of strongly-correlated systems.

The method which is presented here, and the results that it provides, undoubtedly suffer from
a number of disadvantages and shortcomings.  Among those, we can mention that:
\begin{enumerate}
\item it is restricted to one-dimensional, Bethe Ansatz integrable models;
\item it is restricted to the simplest known models, excluding all `nested' systems;
\item it only applies to zero temperature DSFs;
\item it can only treat finite systems;
\item it does not provide an exact, closed-form analytical expression, but only
a numerical result.
\end{enumerate}
Restriction 1 is insurmountable:  the whole ABACUS edifice begins and ends with the Bethe Ansatz.  At best,
what will be achievable is an extension of the framework to provide a 
description of correlations in perturbed integrable models.  Restriction 2 is probably temporary,
and exists simply because it is not yet known whether economical expressions for form factors
of local operators in nested systems exist or can be realistically found.  It is also questionable
whether restriction 3 is temporary or not:  on the one hand, Bethe Ansatz can deal with
finite temperatures quite straightforwardly for equilibrium expectation values.  On the other hand,
the problem of finite temperature Gibbs traces opens up a Pandora's box of complications within the ABACUS logic,
and it is not yet clear whether these can be overcome.  
Restriction 4 comes from the fact that the eigenstates dealt with must be given by finite numbers of
rapidities, and that the Hilbert space must have a measure of finiteness so that summing over the
important intermediate states can be done in a finite time.  It is thus not possible for ABACUS
to give results in the strict infinite-size limit.  The size dependence of the results is a complex
affair, discussed in a later section.  Restriction 5 is the most severe.  ABACUS
is and will remain numerical;  whether simple exact closed-form expressions even exist for general DSFs
is itself more than debatable at present.  There are very good mathematical reasons to believe 
that such expressions cannot be found, except in various simplifying circumstances and/or limits.

For completeness, and in order to relieve the discouragement of the reader, let us now list the
advantages of ABACUS:
\begin{enumerate}
\item it works for very important `canonical' strongly-correlated models such as Heisenberg 
chains and the Lieb-Liniger model;
\item its implementation is relatively universal to the set of integrable models;
\item it provides extremely accurate results for large systems;
\item its accuracy is to a large degree energy and momentum independent.
\end{enumerate}
Advantage 1 is not strictly an advantage, but is worth emphasizing anyway:  the models treated
are strongly-interacting systems which 
{\it do} have experimental applicability, and an important role as beacons of reference for 
refining and fine-tuning alternative approaches.  Advantage 2 comes from the fact that in the
field of integrability, if one knows how to perform a `trick' in one model, one can often
apply the same trick to most if not all other integrable models.  ABACUS is built from the
start with the intension of exploiting this `universality' within the Bethe Ansatz.
Advantage 3 is the most prominent:  the accuracy level can be assessed with sum rules, and
saturation levels beyond $99$\% are achievable for systems with many hundreds of particles,
even in the more difficult limits.  
Relaxing this requirement ever so slightly, say to $90$\%, one can go to systems with thousands of particles.
In any case, the system sizes achievable are comfortably much, much beyond anything which will ever
be realistically achievable using exact diagonalization.  Moreover, the system sizes attainable are also high enough
that finite-size effects are drastically reduced, except perhaps in the vicinity of excitation thresholds,
where the correspondence to singularities which develop in the infinite-size limit seems difficult to obtain.
The fact that large systems can be treated is a consequence of the amount of preliminary information
which is provided by integrability in the first place, and which can be built directly into the algorithm.
The fact that ABACUS can give extremely accurate results for finite systems can itself be seen as an
advantage, when applications to nano/mesoscopic quantum systems are considered.
Advantage 4 is an interesting one, since many other methods (such as bosonization and conformal field theory)
are only valid at low energies.  The common lore is that only `universal' low-energy features should be
computed by theorists, since these are the only ones which are unambiguously observable in experiments.
A more careful assessment however teaches that such low-energy limits are precisely {\it not} accessible
experimentally, and that reliable data read-out has to be done at higher energy scales.  A less misleading
statement about `universality' is thus that the extrapolation via scaling of the low-energy, universal result
(which is not itself consistently observable) to somewhat higher energy scales (which are) can provide
a good fit.  However, the limitations of `universal' approaches quickly becomes severe:  in the case of
spin chains, for example, inelastic neutron scattering provides data over the full Brillouin zone, showing
clearly the band curvature and interaction effects which cannot be quantified using bosonization.
ABACUS shows its strength here by being able to quantify these effects:  it can compute the DSF at any 
energy scale, giving very precise data for any region of the momentum/energy plane.  Since the
Bethe Ansatz used by ABACUS gives exact wavefunctions irrespective of their energy, the accuracy of the data becomes
more or less energy independent (since more types of excited states can live at higher energies, the
accuracy there becomes limited by the restricted number of higher excited states that can
be constructed;  there thus remains an energy dependence to the accuracy, but it is weak).
This feature of ABACUS makes it, at least in the mind of the author, the most appropriate method
to give {\it quantitative} predictions for experiments.

Let us now get down to the specifics.  The bulk of the paper is organized as follows, mirroring
the three elements of the above `wish list'.  Section \ref{sec:Classification} provides an
extensive discussion of the conventions and terminology used in the 
classification and representation of eigenstates within ABACUS.
Section \ref{sec:Constructing} then discusses how individual eigenstates and their
norm are obtained, and how form factors of local operators are computed.  Section \ref{sec:Scanning}
describes how the Hilbert space of intermediate states is scanned in order to achieve optimal
results for the DSF.  
Section \ref{sec:Results} provides some example results obtained from ABACUS, with a discussion
of some of their specific features.  
Finally, section \ref{sec:Discussion} presents a discussion of some generic
features of the approach, and the paper ends with conclusions and perspectives for the future.

\section{Classification and representation of eigenstates}
\label{sec:Classification}
This section aims at providing a representation of eigenstates of Bethe Ansatz integrable
models which is tailor-made for the purposes of calculating correlation functions.  Such
a representation will rely only on very generic features of Bethe eigenstates, so the
discussion will not mention any model in particular (except to highlight exceptional aspects
of specific cases).  

Our integrable model $H$ will have a Fock space ${\cal F}$ of dimensionality $\mbox{dim}({\cal F})$,
which can be finite (for {\it e.g.} finite lattice models) of infinite (for {\it e.g.}
continuum models with no UV cutoff).
Typically, we will be able to easily separate this Fock space into a tensor product of spaces 
${\cal F}_M$ with fixed number $M$ of `particles', with $M$ being the charge of one of the
trivially conserved quantities 
\footnote{For example, for Heisenberg spin chains, where the $\hat z$ component of the total
magnetization $S^z_{tot}$ is conserved, $M$ would be the number of overturned spins starting from
a fully ferromagnetic reference state.}.

\subsection{String basis; configurations}
The Bethe Ansatz gives us explicit wavefunctions for eigenstates of $H$ in a fixed-charge
subspace ${\cal F}_M$, parametrized in terms of $M$ rapidities $\lambda_a$, $a = 1, ..., M$.
In all generality, the rapidities of a given Bethe Ansatz solvable model live in the
complex plane, and this introduces immense complications in the attempt to prove that the basis
of Bethe eigenstates is complete.  Under broad and reasonable assumptions, however, the common
lore is that this basis {\it is} complete \cite{BetheZP71}, and that states can be understood and classified in terms of 
collections of (possibly deformed) generalized `string' patterns
\footnote{This is correct as long as a degree of flexibility is maintained in 
what we mean with a wavefunction of Bethe Ansatz form.  An example that comes to mind is the Heisenberg spin
chain at roots of unity \cite{FabriciusJSP103,FabriciusJSP104}, where a complete set of states includes wavefunctions which can
be viewed as given by the {\it derivative} of a Bethe Ansatz.}.
For a given model, let us call $N_s$ the total number of possible
string patterns.  We let the label $j$ run through the values $1, ..., N_s$, with increasing $j$
representing strings of increasing charge.  In what follows, we call $j$ the string {\it level}.
A given eigenstate will have specific numbers
$M_j$ of strings of level $j$.  We call a set $\{ M_j \}$, $j = 1,...,N_s$ a {\it base}.  
The total charge $M = \sum_j l_j M_j$ where $l_j$ is the charge of a string of level $j$ is simply called
the {\it charge} of a base, and we sometimes write the charge of a base as a subscript 
({\it i.e.} $\{ M_j \}_M$).  The total number of strings, $M_{st} = \sum_j M_j \leq M$ is
called the {\it string charge} of a base.  

Since two-particle scattering preserves strings, the space ${\cal F}_M$ further subdivides
into subspaces ${\cal F}_{\{ M_j \}_M}$ having fixed charge-$M$ bases.  In other words, each
partitioning of $M$ into a charge-$M$ base $\{ M_j \}_M$ generates an independent Hilbert
space.  

The exponential form of the Bethe equations is a set of transcendental equations for the
rapidities.  When written in logarithmic form, these equations can be interpreted as a mapping between 
the set of rapidities $\{ \lambda^j_{\alpha} \}$ and a set of quantum numbers $\{ I^j_{\alpha} \}$
with $j = 1, ..., N_s$ and $\alpha = 1, ..., M_j$ at each level,
\begin{equation}
N \theta^j_{kin} (\lambda^j_{\alpha}) + \sum_{k=1}^{N_s} \sum_{\alpha = 1}^{M_k} \theta^{jk}_{scat} (\lambda^j_{\alpha}, \lambda^k_{\beta})
= 2\pi I^j_{\alpha}
\label{eq:Bethe_Gaudin_Takahashi}
\end{equation}
where $\theta^j_{kin}$ is a kinetic function and $\theta^{jk}_{scat}$ is a scattering phase shift kernel.
Since these equations make use of the string hypothesis (of perfectly undeformed strings whose
deviations are assumed exponentially small in system size), thus 
reducing the number of unknown parameters from the charge to the string charge, it
seems historically appropriate \cite{GaudinPRL26,TakahashiPTP46,TakahashiPTP48}
to call them the Bethe-Gaudin-Takahashi (BGT) equations, reserving the name Bethe equations 
to the original full set (for a number of rapidity parameters equal to the charge) 
of coupled equations without any string hypothesis made.  Although the rapidities are 
all interdependent, the quantum numbers are independent of one another, modulo applying a simple
generalized Pauli principle stating that no two quantum numbers at the same level are allowed
to coincide, {\it i.e.} for $\alpha \neq \beta \in [1, M_j]$, we must have
$I^j_{\alpha} \neq I^j_{\beta}$, $\forall j$.  
The quantum numbers thus provide us with the appropriate labelling of the eigenstates.
We thus write $|\{ I \} \rangle$ or $|\{ \lambda \} \rangle$ for our eigenstates, with the
understanding that $\{ \lambda \}$ are obtained from solving the Bethe-Gaudin-Takahashi equations with 
the choice $\{ I \}$.

The dimensionality of a given ${\cal F}_{\{ M_j \}_M}$ subspace coincides with the
number of allowable choices of quantum numbers, since these obey a generalized Pauli principle \footnote{
In Bethe Ansatz as in many other things, the devil is in the details:  the generalized Pauli principle is
not strictly valid in its most basic formulation, but can be made so by an appropriate cataloguing of string
states, including (over)deformed strings.}.  These can be counted from the observation
that choosing a base fixes the sets of possible quantum numbers $\{I^j_{\alpha} \}$ which can be made at each level.
In other words, the choice of a base $\{ M_j \}$ unambiguously
defines a set $\{I^j_{\infty} \}$, $j = 1, ..., N_s$ of left- and right-limiting quantum numbers 
$I^j_{\infty,-}$ and $I^j_{\infty,+}$ at each level:
any eigenstate in this base must have quantum numbers individually obeying the limiting inequalities
$I^j_{\alpha} \in [-I^j_{\infty,-}, I^j_{\infty,+}]$, $\alpha = 1, ..., M_j$, $j = 1, ..., N_s$ \footnote{While in most cases 
the left- and right-limiting quantum numbers coincide, 
this is not always true.  The simplest example of such a case is 
the isotropic Heisenberg chain, for which one should take $I^1_{\infty,+} = I^1_{\infty,-} + 1$ or the converse,
since the associated rapidity then sits on $\infty$, which is not distinguished from $-\infty$.
Such extra states also occur for example in the gapped $S = 1/2$ antiferromagnet, and in the gapless $XXZ$ chain at roots of unity.  
The states obtained from these exceptional solutions to the Bethe equations are associated to certain discrete
symmetries of the model (the global $SU(2)$ symmetry in the case of the isotropic chain, and the loop algebra
symmetry for the gapless chain at roots of unity), and can be treated in the ABACUS logic  
by for example introducing additional levels.}.
The dimensionality of this subspace is thus
\begin{equation}
\mbox{dim}({\cal F}_{\{ M_j \}_M}) 
= \prod_{j=1}^{N_s} \left( \begin{array}{c} I^j_{\infty,+} + I^j_{\infty,-} + 1 \\ M_j \end{array} \right).
\end{equation}

We call a set of quantum numbers $\{I^j_{\alpha} \}$ fulfilling the generalized Pauli principle 
and the limiting inequalities a {\it configuration}.  A unique quantum number configuration is 
thus associated with each individual eigenstate via the Bethe-Gaudin-Takahashi equations.
One subtlety is that not all configurations lead to sensible string structures.  If a set of
quantum numbers $\{ I^j_{\alpha} \}$ is parity-invariant, then its associated rapidities will
also be parity-invariant.  If quantum numbers $I^{j_1}_{\alpha}$ and $I^{j_2}_{\beta}$ 
for different levels $j_1$ and $j_2$ within such a set are then simultaneously zero, 
both string centers $\lambda^{j_1}_{\alpha}$ and $\lambda^{j_2}_{\beta}$ vanish according to the BGT equations.  If 
the string charges are such that $l_{j_1} = l_{j_2} ~\mbox{mod}~ 2$, then the string hypothesis would predict coinciding
rapidities modulo exponentially small deviations ({\it e.g.} an overlapping $3$- and $5$-string
on zero rapidity leads to 3 pairwise superpositions).  Since the string hypothesis patently fails in such
cases, we call these states {\it inadmissible}, the usual states being {\it admissible}.  
Inadmissible states do give proper wavefunctions involving (strongly) deviated strings, 
but the original Bethe equations must be solved.  ABACUS ignores inadmissible states in the current
implementation, since the numbers of inadmissible states necessarily are suppressed in a $1/N$ fashion as compared
to admissible ones, and in any case their contribution to correlation functions can in some cases be shown to vanish identically \cite{HagemansJPA40}.

For the calculation of zero-temperature correlation functions as performed by ABACUS, 
the most important eigenstates are those for which $M_1 \lesssim M$ and $M_j \sim O(1)$ for $j > 1$,
in other words eigenstates for which only a handful of `higher' strings are present, and most
rapidities sit at level $j = 1$.

\subsubsection{Representation of configurations in ABACUS}
\begin{figure}
\begin{center}
\includegraphics[width=12cm]{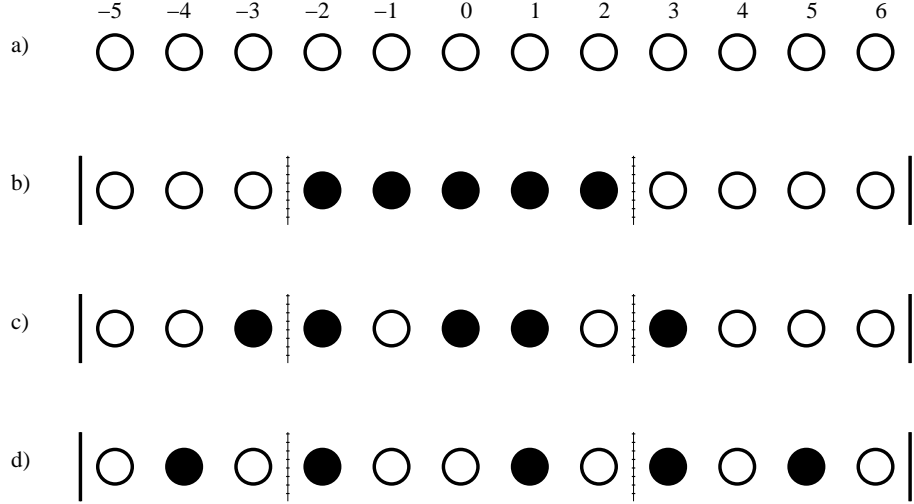}
\end{center}
\caption{Representation of quantum numbers at level $j = 1$.  
In a), the set of possible quantum numbers at this level for a given
base is shown as open circles.  In this case, we must fill in $M_1 = 5$ of the possibilities defined by 
the limiting quantum numbers $I^1_{\infty,-} = 5$ and $I^1_{\infty,+} = 6$.  b) shows the lowest-energy
configuration.  c) shows a relatively low-energy excited configuration with two particle-hole pairs (four excitations) at this level, 
with the particles and holes staying close to the Fermi boundary.  Finally, 
d) gives an example of a high-energy configuration with three particle-hole pairs (six excitations),
with holes going further in and particles further out away from the Fermi boundary.}
\label{fig:Quantum_Numbers}
\end{figure}

Let us now describe how a configuration is represented and labeled in ABACUS.
Let us start by considering level $j = 1$, the level at which we will make use of the most complicated representation.  
The $M_1$ quantum numbers of this level are to be chosen within the set of quantum numbers 
$-I^1_{\infty,-}, -I^1_{\infty,-} + 1, ..., I^1_{\infty,+} - 1, I^1_{\infty,+}$.
We represent the possible choices by open circles, as per a) in Figure \ref{fig:Quantum_Numbers}.
\footnote{Such pictorial representations of quantum numbers and of their movement when scanning through the set of
eigenstates are reminiscent of a traditional abacus, and suggested the name of the library.}
In the same figure, b) represents the lowest-energy configuration, while c) and d) represent
higher-energy configurations.  The fact that b) represents the lowest-energy configuration
comes by design of the rapidity parametrization, and can be ensured at all levels, for all models.

We often use the term `particle' to denote a given occupied quantum number, representing a
string in the eigenstate.  
In most cases, we call `excitations' of a configuration the holes within the base Fermi set (this being defined as the
lowest-energy configuration for a given base), and the
particles outside of it.  Thus, configurations b), c), d) in Figure \ref{fig:Quantum_Numbers} each have 5
particles, but example b) has 0 excitations, while
c) has 4, and d) has six (this is not a strict rule:  a simple exception are the two-spinon
states of Heisenberg chains in zero field, where the spinons are dispersive `holes' \cite{FaddeevPLA85}, and the 
`particles' living out of the base Fermi set are fixed (only one quantum number slot is available);  since the positions of
the two holes completely specify the state, we then say that such states contain two excitations).  
In circumstances where we need to be more specific, we will distinguish between `dispersive' and `non-dispersive'
excitations.  The latter have only one possible choice for their quantum number, and therefore are not part of 
an excitation family;  the former, on the other hand, have many quantum number possibilities, and in most cases become true
excitation branches in the infinite-size limit.  

\begin{figure}
\begin{center}
\includegraphics[width=12cm]{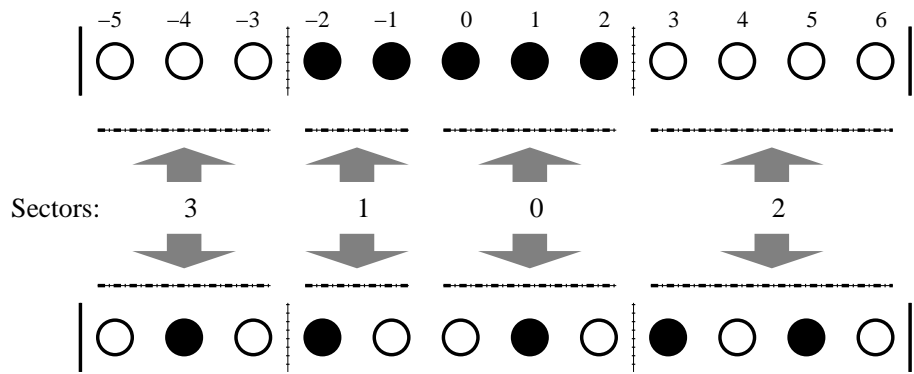}
\end{center}
\caption{The four sectors of excitations at level $j=1$.  
Sectors 0 and 1 are respectively
the right- and left-hand sides of the base Fermi set, while 2 and 3 are the 
right- and left-outsides of the base Fermi set.  The excitation type for the upper
configuration is $0000$.  The excitation type for the lower configuration is $1212$ (see main text).}
\label{fig:Quantum_Numbers_Sectors}
\end{figure}

\paragraph{Sectors.}
Excitations formed by a number of particle-hole pairs can be further separated into different
classes, depending on which sides of the base Fermi set they live on.  We first define four {\it sectors}
as illustrated in Figure \ref{fig:Quantum_Numbers_Sectors}.  
Sector $0$ contains the quantum
numbers obeying $I \geq 0$, $I < I_{F} \equiv \frac{M_1}{2}$ (right-hand side of the base Fermi set,
including zero).  Sector 1 contains those obeying $I < 0, I > -I_{F}$ (left-hand side of
base Fermi set).  Sector 2 contains the quantum numbers for which
$I > I_{F}$ and $I \leq I^1_{\infty,+}$, while sector 3 contains those obeying
$I < -I_{F}$ and $I \geq -I^1_{\infty,-}$.  
The numbers of excitations $e_i^1$, $i = 0, ..., 3$ at level $j=1$, combined and written out as $e^1_3 e^1_2 e^1_1 e^1_0$ 
is called the {\it level $1$ type}
\footnote{We assume that ABACUS will not need to construct states with extremely many
excitations, and we therefore restrict the $e^j_i$ to $0 \leq e^j_i \leq 9$, making the above 
labelling unambiguous.}.  
The fact that excitations come from particle-hole pairs means that at level $j = 1$, 
we have the identity $e_0^1 + e_1^1 = e_2^1 + e_3^1$.  Despite this slight redundancy, 
we still label the type as described above, since it then allows to visualize the overall quantum
number excitation pattern of a state with a quick glance.  Thus, in Figure \ref{fig:Quantum_Numbers},
the configurations b, c and d are respectively of level $1$ type $0000$, $1111$ and $1212$
\footnote{For the specific case of the one-dimensional Bose gas, a different convention is used, 
in which sectors 1/3 and 2/0 are permuted.}.

\begin{figure}
\begin{center}
\includegraphics[width=12cm]{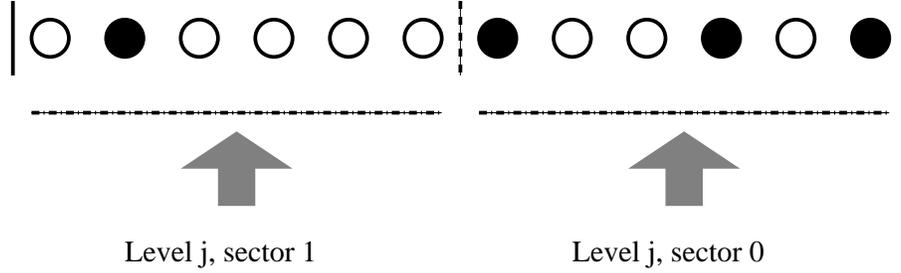}
\end{center}
\caption{The two sectors for levels $j > 1$.  The level $j$ type for this configuration 
is $13$.}
\label{fig:Quantum_Numbers_j}
\end{figure}

Levels $j > 1$ use a slightly less complicated representation.  As mentioned before, 
in practice ABACUS only needs to construct states for which $M_j \sim O(1)$ for $j > 1$, and we therefore
use only two sectors for these levels.  As illustrated in Figure \ref{fig:Quantum_Numbers_j}, 
sector $0$ at level $j$ contains allowable quantum
numbers $I^j \geq 0$, while sector $1$ contains $I^j < 0$.  The {\it level $j$ type} is defined similarly
to the level 1 type, but is here given by the simple combination $e^j_1 e^j_0$.

The whole set of level $j$ types for $j = 1, ..., N_s$ defines the {\it type} of the eigenstate.

\paragraph{Mapping to Young tableaux.}
Within a given sector at a given level, we map the configuration of occupied and unoccupied quantum numbers
onto a Young tableau.  This is done differently in each of the four sectors of level 1, and in
the two sectors of each level $j > 1$.

\begin{figure}
\begin{center}
\includegraphics[width=12cm]{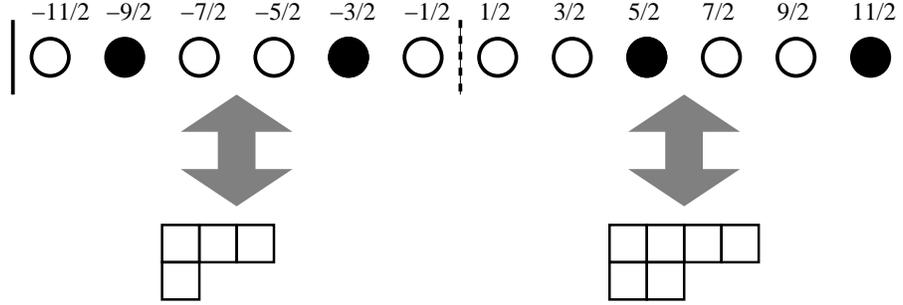}
\end{center}
\caption{Mapping quantum number configurations in a sector to a Young tableau,
here for the two sectors at a level $j > 1$.  }
\label{fig:Quantum_Numbers_Young_j}
\end{figure}

It is simplest to illustrate the general idea using a level $j > 1$.  This is done in
Figure \ref{fig:Quantum_Numbers_Young_j}, where a type $22$ configuration is considered.  
Sector $0$ of that level has $2$ excitations; there are six allowable quantum numbers.
The lowest energy configuration of this type would have the two quantum numbers 
$1/2$ and $3/2$ occupied.  This would be mapped to an empty tableau.  Instead, in
the configuration shown, the rightmost excitation sits on $11/2$, four slots higher than
its quantum number $3/2$ in the lowest configuration.  This displacement is represented
by putting $4$ boxes in the first row of the tableau.  Similarly, the second-rightmost
excitation sits on $5/2$, two slots higher than its original $1/2$.  We therefore put
two boxes in the second row of the tableau.  
In sector $1$, we have two excitations, and we use the same logic but inverted right-to-left.
the first row of the tableau in sector 1 thus represents the left-displacement of the
excitation which sits left-most in the lowest configuration.   It should be more or less
immediately clear to the reader that the generalized Pauli principle means that the
tableaux that are thus obtained obey the Young tableaux rules.  The tableaux in a given sector
can also be read in a dual manner:  if the row lengths represent {\it e.g.} the right-displacements of 
a set of particles/holes, then the column lengths represent the left-displacements of the 
associated holes/particles.  

At level $j = 1$, we need four tableaux.  The sector $0$ tableau at level $1$ represents
the inward left-displacements (towards the middle) of the holes on the right of the base Fermi set.
The sector $1$ tableau similarly represents the right-displacements of the holes on the
left of the base Fermi set.  Sectors $2$ and $3$ represent respectively the right-(left-)diplacements
of the excitations to the right (left) of the base Fermi set, similarly to the logic used
in levels $j > 1$.

For a given sector $i$ at a given level $j$, the number of rows $n_r$ in the corresponding tableau
is equal to the number of excitations $e_i^j$ in this sector.  The number of columns $n_c$ is given
by the maximal possible displacement of the `highest' excitation, which is simply the
number of available quantum numbers $n_i^j$ in this sector minus the number of excitations.
There are thus $d_i^j = \left( \begin{array}{c} n_i^j \\ e_i^j \end{array} \right) \equiv B_{n_i^j, e_i^j}$ different
allowable tableaux (in other words, configurations) in this sector at this level.  

\begin{figure}
\begin{center}
\includegraphics[width=12cm]{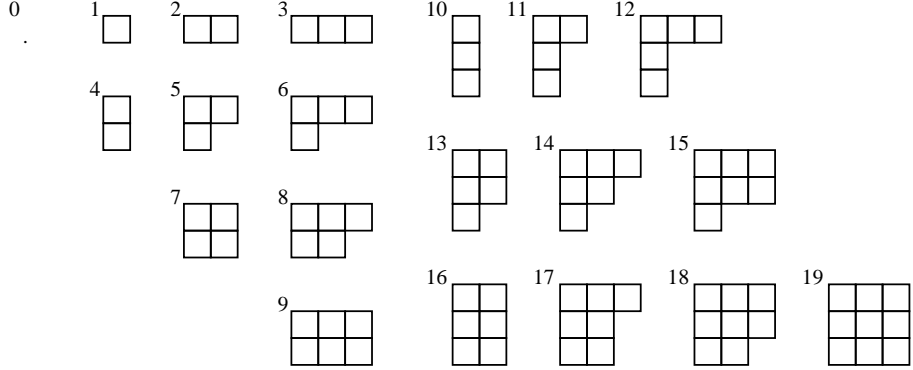}
\end{center}
\caption{Young tableaux and their identification numbers.  Tableaux $0$-$9$ form the set
of all ten tableaux with $(n_r, n_c) = (2, 3)$.  Tableaux $0$-$19$ form the set of twenty tableaux
with $(n_r, n_c) = (3, 3)$.}
\label{fig:Young_IDs}
\end{figure}

We number Young tableaux in the following manner:  the empty tableau is given $id = 0$.
We add boxes on the first row until the row is full, increasing the $id$ by one for each
added box.  When we reach the maximal length, the next tableau is the one with a single
box on the second row (and thus also on the first row).  
We then add boxes again on the first row up to the maximal length.
The next tableau is then the one with two boxes on the second row.  We proceed like
this until both the first and second rows are full.  The next tableau is then the one
with one box on the third row (and thus also on the second and first ones).  This is
done until the full tableau is obtained.  An example of this is given in Figure
\ref{fig:Young_IDs}, where we list all the $(n_r, n_c) = (3,3)$ tableaux and their identification
numbers.  It is simple to write a recursive algorithm providing the mapping to/from
Young tableaux from/to identification numbers.  

\paragraph{Eigenstate labelling in ABACUS.}
We now have all the needed elements to describe the labelling of eigenstates within ABACUS.  
An eigenstate is uniquely identified by a configuration.  A configuration is uniquely
defined by three integers:  its {\sf base\_id}, its offset {\sf type\_id}, and its {\sf offset\_id}.

The {\sf base\_id} is a {\sf long long integer} specifying the set $\{ M_j \}$.  In general, we have
$M_1 \simeq M \gg 1$, with $M_j \sim O(1)$ for $j > 1$.  We thus write
\[
{\sf base\_id} = [\#\#]_{M_{N_s}} [\#\#]_{M_{N_s-1}}...[\#\#]_{M_{3}} [\#\#]_{M_{2}} [\#\#\#\#\#]_{M_1},
\]
in which $[\#\#]_{M_j}$ signifies two integers giving $M_j$, {\it e.g.} $03$ or $11$ or other two-digit non-negative number
less than $100$.  At level $1$, in order to permit the labelling of states in large systems, 
we allow for a five-digit specification.  The numbers should be
read from {\it right to left}, since leading $0$'s are scrapped by file or standard output.  A 
{\sf type\_id} of $1040200293$ therefore means that the base is given by
$M_1 = 293$, $M_2 = 2$, $M_3 = 4$, $M_4 = 1$ and $M_j = 0$, $j > 4$.   

The {\sf type\_id} of the offset of a configuration is represented as an integer 
\[
{\sf type\_id} = e_1^{N_s} e_0^{N_s} e_1^{N_s-1}e_0^{N_s-1}.... e_1^3 e_0^3 e_1^2 e_0^2 e_3^1 e_2^1 e_1^1 e_0^1
\]
(N.B.:  the numbers should again be read {\it from right to left}).  This is also represented as
a {\sf long long integer} in ABACUS, since in practice the code never puts more than nine
excitations in a sector (allowing the {\sf type\_id} to be interpreted one character at a time), 
and uses only the lowest set of levels (meaning that the limitations on the length of integers 
do not affect the actual calculations performed).  

Finally, the {\sf offset\_id} is defined as follows.  Recall that $d_i^j$ is the total number
of tableaux of sector $i$ at level $j$.  The tableaux identification number $t_i^j$ for this sector
at this level thus runs through the values $0, 1, ..., d_i^j - 1$.  Given a set of tableaux identification
numbers $\{ t_i^j \}$ for each sector at each level, we thus define
\[
{\sf offset\_id} = \sum_{j=1}^{N_s} \sum_{i=0}^3 t_i^j \prod_{i'=0}^{i-1} d_{i'}^j
\prod_{j'=1}^{j-1} \prod_{i''=0}^3 d_{i''}^{j'}
\]
as the single integer defining all tableaux identification numbers.  
In other words, we order the $(s, l)$ sector/level 
pairs as 
\begin{equation*}
(0, 0), (1, 0), (2, 0), (3, 0), (0, 1), (1, 1), (0, 2), (1, 2), ..., (0, N_s -1), (1, N_s -1),
\end{equation*}
and the {\sf offset\_id} is obtained by multiplying each tableau id by
the product of the number of possible tableaux at each lower sector/level pair.

We therefore have defined a mapping between three integers, and the whole set of quantum numbers of
an eigenstate:
\[
({\sf base\_id}, {\sf type\_id}, {\sf offset\_id}) \Leftrightarrow \{ I^j_{\alpha} \}
\]
where the string content and excitation pattern are clearly ({\it i.e.} humanly) legible 
from the first and second integers respectively.  These numbers are used in all functions used
by ABACUS, and in the data files that it produces.
%\footnote{All data structures are defined as {\sf public}, and can therefore be user-manipulated.
%Thus, the quantum numbers themselves can be user-modified without references to the identification
%numbers.}.  % FOR USER GUIDE !

\section{Constructing individual eigensates and calculating their form factors}
\label{sec:Constructing}
\subsection{Solving the Bethe equations}
\label{subsec:Solving}
As described above, the Bethe Ansatz provides an economical pathway for obtaining eigenfunctions
by replacing the diagonalization of the Hamiltonian by the process of solving some sets of
nonlinear coupled equations for one eigenstate at a time.  The overall difficulty of solving the
Bethe equations varies greatly from one model to the other.  Also, within one model, 
the solution to the Bethe equations are easy to find for some of the eigenstates, while for some
other eigenstates this might represent a very difficult challenge.

The simplest situation is when all solutions to all Bethe equations are to be found in terms
of real, finite rapidities.  This is the case for example for the one-dimensional Bose gas
(Lieb-Liniger model) in the repulsive regime:  the proof is simple, and relies on the convexity of the Yang-Yang
action associated to the Bethe equations (\ref{eq:Bethe_Gaudin_Takahashi}) on the field of real numbers
(\cite{YangJMP10}, see also details in \cite{KorepinBOOK}).  For Heisenberg spin chains however, the corresponding Yang-Yang action
is not convex, and complex solutions exist.  It is a well-known fact that there is no, and cannot
be, a good numerical method for solving coupled nonlinear systems of equations, so we have to 
input as much preliminary knowledge as possible in the procedure.  The string hypothesis conveniently 
provides such preliminary information.

In order to solve the Bethe-Gaudin-Takahashi equations, ABACUS mixes different methods.  The simplest method,
used in the initial stages of the solution process, consists in performing some simple iterations.
Defining the set $\{ \lambda^j_{\alpha, (i)} \}$ as the set of rapidities at iteration number $i$, we can
proceed as follows.  First, the initial set is defined as the solution to the decoupled
conditions $N \theta_{kin}^j (\lambda^j_{\alpha, (0)}) = 2\pi I^j_{\alpha}$.  Subsequent sets
are defined from the recursion relation $N \theta_{kin}^j (\lambda^j_{\alpha, (i+1)})
=  2\pi I^j_{\alpha} - \sum_{k=1}^{N_s} \sum_{\beta = 1}^{M_k} \theta^{jk}_{scat} (\lambda^j_{\alpha, (i)}, \lambda^k_{\beta, (i)})$,
possibly using some damping to help stabilize the process.  Such simple iterations have the
advantage that they often work, in the sense that some measure of convergence is
achieved.  Each iterative step is also rather quick, involving order $M^2$ operations.  
However, convergence is slow, especially when the solution is approached, and it is therefore desirable to accelerate the
algorithm in various ways.

One way to converge more quickly is to track the changes of the rapidities over a certain number of
iterations.  This gives a flow pattern to each rapidity:  if this flow is sufficiently regular,
it can then be extrapolated to an infinite number of iterations in order to generate a new
set of rapidities.  In practice, this procedure of combining iterations (say four or five) 
and extrapolations substantially accelerates convergence as compared to doing only simple iterations.  

A more elegant approach becomes available once the iterations have come sufficiently close to the
true solution:  the matrix Newton method.  This has the advantage of converging quadratically,
but has the disadvantages that 1) it is more computationally costly per iteration step, involving order $M^3$ operations,
and 2) it is not always stable if the starting set of rapidities is too far from the actual solution
(this instability is a possibility if the Yang-Yang action is not convex).
Various run-time checks need to be made to see if this procedure is stable.  One advantage of the matrix Newton
method is that the Jacobian matrix which is needed, given by the matrix of derivatives of the left-hand
side of (\ref{eq:Bethe_Gaudin_Takahashi}), is precisely the Gaudin matrix which is needed to compute the norm of the
eigenstate (discussed in the next subsection), which also provides a slight acceleration of the algorithm.

ABACUS considers the Bethe equations as solved provided a further iterative or Newton step 
gives a new set of rapidities for which the sum of square differences with the previous set is below
a fixed convergence precision threshold, the latter being set to a fixed value somewhat 
larger than the available numerical accuracy (machine epsilon).  In the case of purely real rapidities,
this is the only convergence criterion applied.  For states with strings, however, we have to be more
careful.  Namely, solutions to the Bethe-Gaudin-Takahashi equations do not always represent proper solutions
to the corresponding Bethe equations, since the string hypothesis might not be verified to sufficident accuracy for that specific
eigenstate.  In spin chains, this occurs in many circumstances, including for eigenstates with strings having a very large rapidity,
or when the magnetic field is near zero
(for a more extensive discussion of deviated strings, see \cite{HagemansJPA40} and references therein).  
For eigenstates with string states, ABACUS therefore also computes the first-order string
deviations by substituting the solution of the Bethe-Gaudin-Takahashi equations back into the original Bethe
equations, yielding a new iteration for the (now full) set of rapidities, this time including their imaginary parts
explicitly.  If the sum of the string deviations (defined as the difference between the patterns obtained
and pure, undeformed strings) is higher than a given fixed string 
precision threshold, the eigenstate is deemed unusable, and its contribution is excluded from the final result
to prevent corruption of the data.  If these deviations are small enough, the contribution from this state
will however be kept.

\subsection{Calculating norms and form factors}
\label{subsec:Norms_and_FFs}
Once the set of rapidities $\{ \lambda^j_{\alpha}\}$ of an eigenstate is known, it is then a matter
of straightforward number crunching to obtain the norm of the eigenstate.  This is given by the
Gaudin-Korepin formula \cite{GaudinJMP12,KorepinCMP86} in terms of the determinant of the Gaudin matrix, the latter being defined (as mentioned
above) as the derivative matrix of the Bethe equations (in the case of string states, the norm can
be similarly calculated from a reduced Gaudin matrix obtained from the Bethe-Gaudin-Takahashi equations,
see \cite{KirillovJMS40,CauxJSTAT05P09003}).  The determinant itself is then calculated using LU decomposition.

For form factors of local operators in non-nested Bethe Ansatz integrable models, the Algebraic
Bethe Ansatz provides a representation in terms of a matrix determinant, similarly to the norm.
In this case, however, the matrix depends on two sets of rapidities (one for each of the bra and 
ket states), and takes a different form depending on which operator is considered.  The explicit
formulas can be found in the literature.  For the Bose gas, the density operator form factor is given
in \cite{SlavnovTMP79,SlavnovTMP82}.  The field operator form factor was given as a sum of determinants
in \cite{KojimaCMP188}, and simplified to a single determinant in \cite{CauxJSTAT07P01008}.  For Heisenberg  
$S = 1/2$ chains, the $S^z$ and $S^{\pm}$ operator form factors were given in \cite{KitanineNPB554,KitanineNPB567} for the
case where string states exist at most on one side of the bracket.  The general expression, valid when either (or both) of the bra and ket
states have strings, was given in \cite{CauxJSTAT05P09003} \footnote{There is an unfortunate typo in equation (29),
where $L$ should be defined with a minus sign.}.  
Here again, the determinant is calculated by ABACUS using LU decomposition.

\section{Scanning the Hilbert space}
\label{sec:Scanning}
Assuming that eigenstates can be classified, individually labeled and constructed as described
in the previous sections, we now come to the more challenging third item on the introduction's 
`wish list':  the summation over intermediate states.

For a given DSF in a given model, the perfect scanning algorithm will generate eigenstates
in order of numerically decreasing absolute value of form factors.  
This is not trivial to achieve in the models we
consider, since the intuitions we can develop relating to the solutions to the Bethe
equations and the rapidity dependence of the form factors, remain incomplete.  

All is not lost, however.  
Four general but simple guiding principles can be used to define an efficient algorithm to scan the
Hilbert space.  The first principle is that i) for large enough system sizes, form factors
obey an approximate continuity principle, in the sense that modifying an excited state in
a small way ({\it e.g.} moving just one quantum number) does not dramatically change the form factor.  
This principle is ultimately related to the fact that form factors are analytic functions of the rapidities of the eigenstates involved, and
translates to the statement that if we find some excited state
having a large contribution, we can exploit this by concentrating computational
resources on the pool of states in its vicinity.  

The remaining three principles associate in turn to the three integers we use to
label eigenstates.  In general, we observe that form factors of local physical operators
between the ground state and an excited state decrease in numerical value for increasing
complexity of the excited state, by which we explicitly mean that form factors become smaller
if ii) more/higher strings are included in the excited state, iii) more particle-hole excitations
are created in the lowest level and iv) the excitations' quantum numbers are moved further and
further away from the base Fermi set defining the lowest-energy state. 

These principles are not strictly true in all circumstances, and the ABACUS implementation
attempts to `take them with a grain of salt' and probe sufficiently widely to gather all 
important contributions on the one hand, but not spend too much time on irrelevant contributions
on the other.  

\subsection{Climbing the Bethe tree}
\label{subsec:Bethe_Tree}
The best way to visualize the scanning through the Hilbert space for the excited states 
is by using an analogy with another of Bethe's creations, namely a Bethe lattice
\cite{BethePRSLA150}.  
Starting from the lowest-energy state, one can then create particle-hole pairs in the
lowest level, let these particles and holes move away from the Fermi configuration, 
add higher strings, let those disperse, and so on.  

\begin{figure}
\begin{center}
\includegraphics[width=12cm]{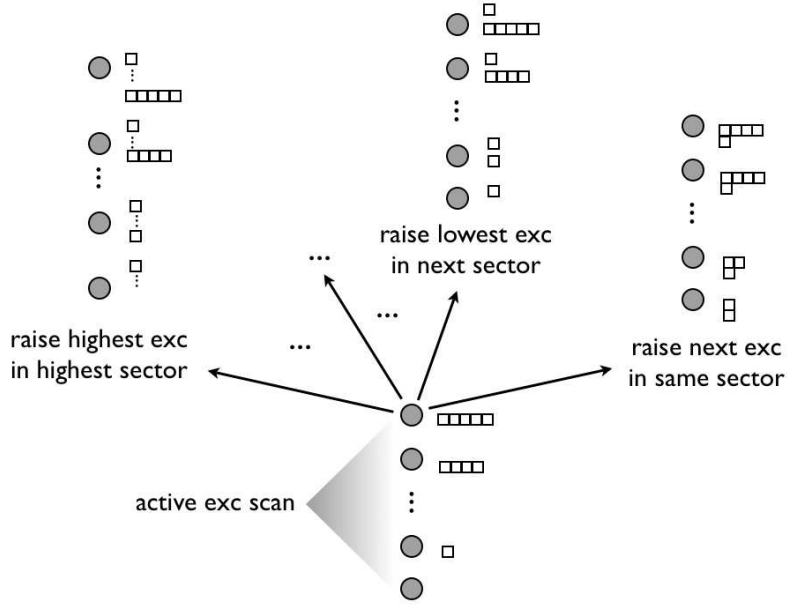}
\end{center}
\caption{Climbing the Bethe tree for a given base.  If an active excitation scan gives 
good enough results (in the sense defined in the text), a whole new generation of scans
is initiated by descending the state (again, as explained in the text) in all possible ways, 
and starting a new active excitation scan for each such descendent.  This procedure is
continued recursively, and continues until good results are exhausted. 
Since each eigenstate is obtained
via a unique such descendency path, the full Hilbert space can be faithfully probed.}
\label{fig:Bethe_tree}
\end{figure}

Explicitly, the first excited state which is constructed is the lowest-energy state of the appropriate subspace,
with the lowest {\sf base\_id} and zero {\sf type\_id} and {\sf offset\_id}.  
A particle-hole pair is then created, the {\sf type\_id} becoming in turn 
$101$, $110$, $1010$ and $1001$.  For each of these four branchings, a scan is made
on the position of innermost (so sector $0$ or $1$) excitation.  If the contributions obtained 
to a chosen sum coming from these states is `large enough' (whose meaning we will make explicit below), 
the next innermost excitation is `raised' by a unit, and a new scan is made over the innermost excitation positions.
The same logic is applied recursively in all sectors of all levels of the base in current use,
as illustrated in figure \ref{fig:Bethe_tree}.  

If a base, through this scheme, has given a large enough contribution, the base is then
`complexified' at different levels, yielding a new family of bases on which the recursive scanning
is also performed in a similar way until all large enough contributions from all bases have been exhausted.
Very importantly, an ABACUS object called {\sf Scanned\_Intervals} maintains a map of the Hilbert space
which shows which regions have been scanned, and what average results have been obtained per scanned region.
This crucial run-time information will be used subsequently, as described in the next subsection.

\subsection{Optimizing the climb}
\label{subsec:Optimizing}
Defining the term `large enough' used in the previous subsection leads us to the description of
the highest-level procedure in ABACUS.  The driving algorithm defines a real quantity called
{\sf running\_threshold} which defines a `large enough' form factor as one whose value exceeds
this threshold.  The initial value of the threshold is chosen such that only relatively few
(of order $\sim N^2$) intermediate states are constructed on a first run of the recursive Bethe
tree climb.  Once this first climb has terminated, a new lower value of {\sf running\_threshold}
is chosen, and a new climb is initiated.  

The {\sf Scanned\_Intervals} object now fulfils its role, by 1) specifying to the scanning
algorithm which parts of the Hilbert space of intermediate states can potentially yield
contributions which are now `large enough' according to the new {\sf running\_threshold}, and
2) which parts have already been scanned and should therefore be avoided to prevent state
double-counting.  

When the second recursive Bethe tree climb terminates, the {\sf running\_threshold} is again
reduced by an appropriate amount, and a third recursive climb is initiated.  This is continued
until a user-specified maximal allowed computing time is reached, before which ABACUS interprets and
saves all the data to disk.  Importantly, the {\sf Scanned\_Intervals} object is also saved,
allowing a calculation which has successfully terminated to be restarted at a later stage,
enabling to `polish' results if required.  A parallel implementation of ABACUS also exists,
based on the same principles.  This whole construction naturally
provides an optimal use of the available computing resources.

\section{Results and their interpretation}
\label{sec:Results}
\begin{figure}
\includegraphics[width=12cm]{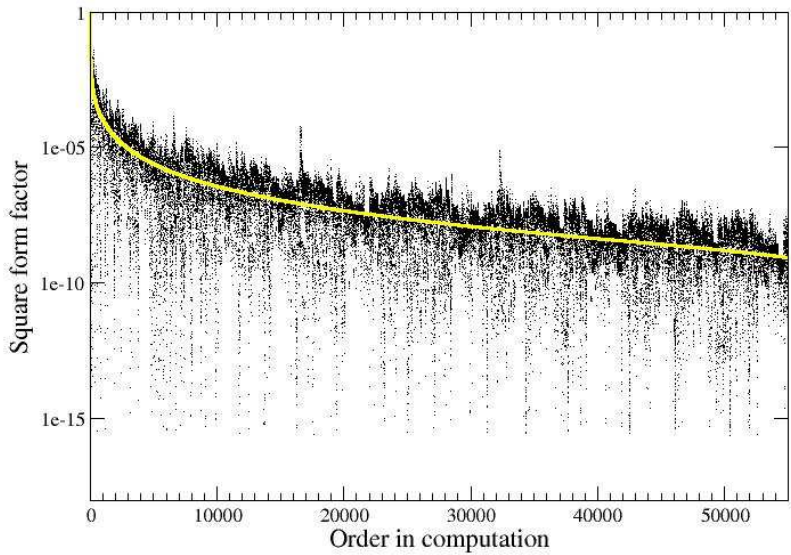}
%\includegraphics[width=12cm]{FFsq_vs_order_D0p6N50M20.eps}
%\psfrag{ABACUS}[][][][]{$|F^-|^2$}
\caption{Numerical value of (squared) form factors obtained during a typical ABACUS run
(here, for the $XXZ$ model at $\Delta = 0.6$ and $S^z_{tot} = 0.1$ for a small system of $N = 50$ sites).
The horizontal axis is simply the ordinality of the computed point.  The continuous (in color:  yellow)
line represents a hypothetically-achieved perfect ordering of form factors in monotonically decreasing
numerical value (an ideal algorithm would simply follow the yellow line).  
While still far from perfect (a factor of approximately 10 in speed could be gained by making the scanning
perfect), ABACUS does follow such ideal lines reasonably closely in most
circumstances.}
\label{fig:FFsq_vs_order}
\end{figure}

At the end of an ABACUS run, a large collection of form factor data becomes available,
composed of energy-momentum-form factor triplets, either in a file with additional state-by-state
information, or conveniently binned into a form factor matrix (for large-scale computations 
producing immense numbers of entries, which could not all be individually saved to disk).
The quality of this raw set of data (see an example in Figure \ref{fig:FFsq_vs_order})
can be measured by exploiting various sum rules, the simplest
of which is to consider the summation over all momenta and integration over all frequencies of equation \ref{eq:DSF}.
The resulting local expectation value $\langle {\cal O}^a{\cal O}^{\bar a} \rangle$ is in most cases
known analytically.  Since expression \ref{zero_T_DSF} is a summation over terms strictly greater than or 
equal to zero, the form factor data set can be summed up and compared to the analytical expectation value.
A sum rule saturation percentage can thus be assigned to each data set produced.  Other sum rules can
occasionally be used, for example the so-called $f$-sum rule which relates the integral over all frequencies
of $\omega S^{a\bar{a}} (k, \omega)$ to a known function of $k$.  This is particularly useful since it
allows to individually evaluate the quality of each separate momentum slice in the data set, which is
beyond the reach of the integrated intensity.  Sum rules are routinely checked by ABACUS:
a summary file is produced, which contains information about the run.  For each base and associated type, 
the number of scanned states is given, together with their contribution to the integrated intensity sum
rule.  Such summary files therefore provide a wealth of information about which types of excitations 
carry significant correlation weight, and which don't.

\begin{figure}
\includegraphics[width=12cm]{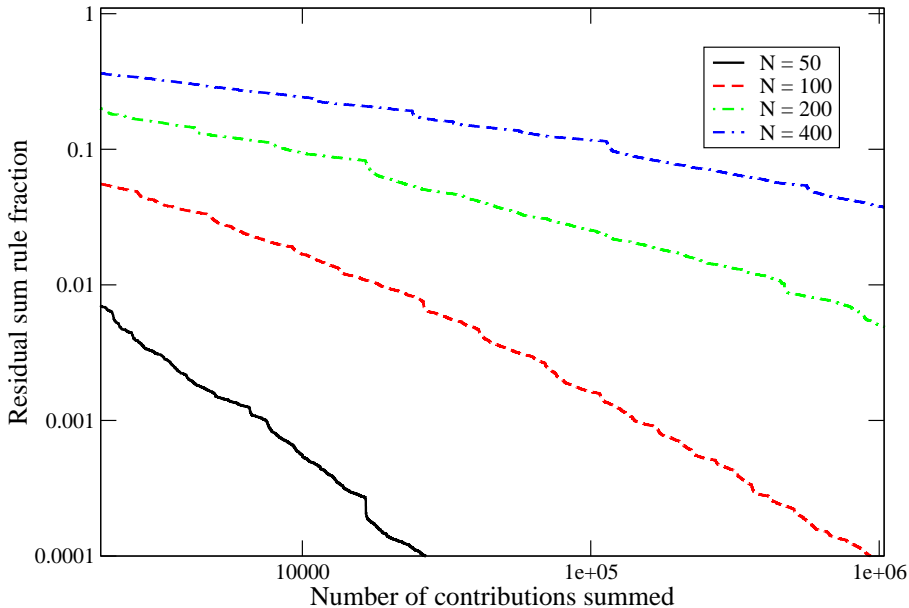}
\caption{Again for the typical example of the $XXZ$ model at $\Delta = 0.6$ and $S^z_{tot} = 0.1$, 
the remainder of total integrated intensity sum rule after summing finite numbers of form factors 
is plotted.  The number of states needed to achieve a required saturation increases more rapidly than
polynomially in system size (as seen by the decreasing slope of the curves for increasing $N$), 
pointing to additional weak system size dependence in the ABACUS operation count exponent associated
to the system size (see text for further discussion).}
\label{fig:SR_vs_order}
\end{figure}

\begin{table}
\begin{center}
\begin{tabular}{|c|c|c|c|c|}
\hline
\% & N = 50 & N = 100 & N = 200 & N = 400 \\
\hline \hline
90\% & 235 (4.9e-12) & 944 (6.8e-26) & 8772 (5.3e-54) & 117373 (3.5e-111)\\
95\% & 385 (8.2e-12) & 2492 (1.8e-25) & 27096 (1.6e-53) & 573466 (1.7e-110) \\
99\% & 1515 (3.2e-11) & 18490 (1.3e-24) & 469120 (2.8e-52) & \\
99.9\% & 7478 (1.6e-10) & 141560 (1.0e-23) & & \\
99.99\% & 26724 (5.7e-10) & 932706 (6.8e-23) & & \\
%90\% & 235 (4.9e-12; 40) & 944 (6.8e-26; 135) & 8772 (5.3e-54; 731) & 117373 (3.5e-111, 5175)\\
%95\% & 385 (8.2e-12; 82) & 2492 (1.8e-25; 386) & 27096 (1.6e-53; 4319) & 573466 (1.7e-110; 34261) \\
%99\% & 1515 (3.2e-11; 386) & 18490 (1.3e-24; 2500) & 469120 (2.8e-52; 16359) & \\
%99.9\% & 7478 (1.6e-10; 1342) & 141560 (1.0e-23; 15366) & & \\
%99.99\% & 26724 (5.7e-10; 2182) & 932706 (6.8e-23; 84047) & & \\
\hline
\end{tabular}
\end{center}
\caption{Number of form factors needed to attain a given sum rule saturation level, for
various values of system size.  This is again for the example of the $XXZ$ model at 
$\Delta = 0.6$ and $S^z_{tot} = 0.1$.  The numbers in parentheses are the Hilbert space
dimensionality fraction represented by the number of states mentioned, illustrating the
efficiency of working in the chosen eigenstates basis.  
%The second number in
%parentheses is the number of contributions needed to achieve the same saturation, for a perfect
%algorithm summing form factors in a monotonically decreasing fashion.  
Missing numbers in the table
could be obtained by simply running ABACUS for longer (all data here is from single-CPU runs).  Although the 
number of states needed to achieve a fixed target sum rule saturation increases more
rapidly than polynomially in $N$ as seen from Figure \ref{fig:FFsq_vs_order}, the Hilbert space fraction needed also decreases.}
\label{table:saturation}
\end{table}

Inevitably, for a given target sum rule saturation, the computation quickly becomes more intensive with 
increasing system size.  The operation count for obtaining an individual state and its form factor scales like
$N^3$.  In practice, the number of states that has to be included in the summation over intermediate
states superficially seems to scale polynomially with $N$ as $N^{\sim 3-5}$ depending on the system and its parameters.
As a rule of thumb, doubling the system size thus requires about two orders of magnitude more 
computational power to achieve the same saturation.  
In fact, the scaling with system size is a complex affair.  
If this scaling was of a well-defined degree, the curves of leftover sum rule weight as a function of number
of contributions summed would have the same
slope (assuming the efficiency of ABACUS to be system size independent).  This is not the case, 
as can be seen from the example in Figure \ref{fig:SR_vs_order} and Table \ref{table:saturation}, 
and there seems to be some additional (possibly logarithmic) system size dependence in the exponent.
This is in fact quite natural: for larger and larger systems, we expect an increasing fraction of the
correlation weight to be carried by states with higher excitation numbers.  The system size scaling is
discussed somewhat more in the discussion section, but a systematic study of it is beyond the scope of
this article.

The finite system dynamical structure factor which ABACUS calculates, as is clear from equation \ref{zero_T_DSF},
is not a smooth function in the energy-momentum plane, since it is simply composed of isolated delta
peaks of various heights, aligned on well-defined lines in momentum but distributed in a very irregular
pattern in energy (due to the adopted periodic boundary conditions on the wavefunctions, momentum is always an 
integer multiple of $2\pi/N$;  energies are on the other hand certainly not equispaced in our interacting models).  
Only in the thermodynamic limit does the DSF become a continuous function;
to obtain this, some form of smoothing of the ABACUS results is necessary.  Since the momentum is already
sitting on a regular lattice, only the energy delta function is smoothened into a gaussian.  This has the
unfortunate effect of blurring some would-be sharp excitation thresholds, but allows to obtain easily
interpretable density plots of the DSF.  The width of the gaussian is chosen so as to be somewhat larger
than the typical 2-excitation energy level spacing, so that the delta functions blend into a smooth function
representing the density of states in this region of the Brillouin zone.  Since this 2-excitation energy level
spacing is typically of order $1/N$, the negative effects of the gaussian quickly become less significant
for bigger systems.  

\section{Discussion}
\label{sec:Discussion}
\subsection{Size dependence of contributions}
The dependence on system size of a given DSF in a given model is very complex.  First of
all, it may take different forms depending on which model parameters are used, and
on which part of the energy-momentum plane one is studying.  Changing system size affects
i) the number of eigenstates in the Hilbert space, ii) the relative number of states of
different bases, iii) the quality of numerical solutions to the Bethe equations, in particular
the relative number of acceptable and unacceptable quality string states that can be
constructed, iv) the form factor of a given identifiable state, and finally 
v) the positional distribution of eigenstates in energy (density of states).

Point i) is trivial:  the dimensionality of the Hilbert space is factorial in system size,
and poses a severe limitation on the attainable size.  
Point ii) is more subtle and interesting.  Namely, the number of $n$-excitation states
scales approximately as (system size)$^n$;  therefore, the number of $2, 3, 4, ...$ excitation
states scales differently, families with more excitations having an increasingly larger
number of members.  For any $n$-excitation state, we can generically
find order (size)$^2$ states with $n + 2$ excitations, and determining how these different bases relatively
contribute to the final answer for large systems is thus an extremely complicated problem.

We can however make some general statements.  Suppose for definiteness that our excitations
only come in pairs (as in {\it e.g.} the Lieb-Liniger model).  For very small size
(few particles), $2$-excitation states will carry a fraction $c_2 \sim 1$ of the sum rules.
As the system size $N$ increases, we will see the $2$-excitation contribution `leak' (as guaranteed by 
the preservation of sum rules)
into the increasingly more numerous $4$, $6$, $8$...-excitation ones.  
$c_2$ will be strictly monotonically decreasing in $N$, 
{\it e.g.} $\frac{\Delta c_2}{\Delta N} < 0$ $\forall N$.  Despite this, the limit
$\lim_{N \rightarrow \infty} c_2$ might remain finite (as is the case for spinons in the zero-field isotropic 
or gapped Heisenberg antiferromagnet;  for the gapless region, it seems reasonable for
$c_j$ for $j$ finite to vanish in the thermodynamic limit, except at zero field).  $c_4$, on the other hand, will not
be monotonic:  it will increase for small $N$ but eventually reach a maximum at some
inflexion value $N_4$, after which it will decrease, {\it i.e.}
$\frac{\Delta c_4}{\Delta N} > 0$, $N < N_4$ but $\frac{\Delta c_4}{\Delta N} < 0$, $N > N_4$.
The limiting value $\lim_{N \rightarrow \infty} c_4$ might also remain finite.  
There will similarly be increasingly large inflexion values $N_6$, $N_8$, ..., $N_4 \ll N_6 \ll N_8....$,
but the determination of these inflexion values is beyond the reach of the current implementation.

This scenario, although more or less inevitable in the mind of the author,
remains conjectural since even the lowest of these inflexion points ($N_4$) seems
to lie above currently achievable system sizes.  In any case, we can expect
the finite-size result for large enough systems 
to closely resemble the infinite-size result (except perhaps
in the immediate vicinity of excitation thresholds):  while each term in the series 
$c_2 + c_4 + c_6 + ...$ might have substantial size dependence, the sum itself ({\it i.e.} the full DSF)
is seen to have much smaller size dependence:  increasing contributions (as system size increases) from
higher excitation numbers tend more or less to compensate the decrease of the lower excitation
number contributions (at least in gapless models), since these two types of contribution resemble each other.
Thus, for example, the contribution of four-spinon states in the
Heisenberg magnet \cite{CauxJSTAT06P12013} looks like a rescaled
two-spinon contribution, and we can expect more or less the same of yet higher
excitation numbers (this, disregarding the different high-energy tails, which
carry very little correlation weight anyway).

\section{Conclusion and perspectives}
Much remains to be done in the general field of dynamical correlation functions of
integrable models, and the reader is referred to \cite{CauxADVPHYS} for an extensive
discussion of this subject.  As far as ABACUS is concerned, various improvements of the
sub-algorithms for solving Bethe-Gaudin-Takahashi equations are possible, as well as further
optimization of the state scanning algorithm.  A better handling of deformed string
eigenstates is also necessary:  this, at the moment, represents the most severe
limitation of the applicability of ABACUS for some DSFs of spin chains in small
magnetic fields, but requires
a rather more elaborate treatment to be properly dealt with.  For certain cases, as for 
example gapped antiferromagnets, the correct state counting itself is not fully known
for arbitrary bases due to difficulties in defining the quantum number limits
(see \cite{CauxJSTAT08P08006} for further discussion of this point).  

The restriction to zero temperature would also need to be overcome, possibly by using some
ideas from the thermodynamic Bethe Ansatz.  On a more theoretical front, representations
of form factors for nested systems such as the Yang permutation model \cite{YangPRL19} or
the Hubbard model \cite{HubbardBOOK} are needed before ABACUS can be used.  
Once these are available however, the `universal' ABACUS logic will be quickly 
implementable for these systems as well.

\paragraph{Data availability}
Requests to the author for calculations of dynamical structure factors for specific models and values of the parameters
are welcome.  An online database of ABACUS computations will also eventually be made available.
Alternately, the {\sf C++} source code can be obtained from the author on request.

\section{Acknowledgements}
Much work went into the development of ABACUS, which was initiated during a collaboration
with J.M. Maillet.  Subsequent versions of the spin chain part of the 
code were checked against an independent implementation by my former student R. Hagemans.
In a similar fashion, the one-d Bose gas part of the code was proof-checked against an independent 
implementation by P. Calabrese.

\end{document}